\documentclass[12pt]{article}

\catcode`\@=11
\@addtoreset{equation}{section}

\global\arraycolsep=2pt
\usepackage{amsbsy,amssymb,latexsym,amsfonts,amsmath}
\usepackage{graphicx,color}

\allowdisplaybreaks

\begin{document}
\begin{flushright}
\parbox{4.2cm}
{IPMU12-0030}
\end{flushright}

\vspace*{0.7cm}

\begin{center}
{\Large \bf Holographic Renormalization of Foliation Preserving Gravity and Trace Anomaly}
\vspace*{1.5cm}\\
{Yu Nakayama}
\end{center}
\vspace*{1.0cm}
\begin{center}
{\it Kavli Institute for the Physics and Mathematics of the Universe, \\ 
Todai Institutes for Advanced Study, 
\\ the University of Tokyo, Kashiwa, Japan, 277-8583 (Kavli IPMU, WPI)}
\vspace{3.8cm}
\end{center}

\begin{abstract} 
From the holographic renormalization group viewpoint, while the scale transformation  plays a primary role in holographic dualities by providing the extra dimension, the special conformal transformation seems to only play a secondary role. We, however, claim that the space-time diffeomorphism is crucially related to the latter. For its demonstration, we study the holographic renormalization group flow of a foliation preserving diffeomorphic theory of gravity (a.k.a. space-time flipped Horava gravity). We find that the dual field theory, if any, is only scale invariant but not conformal invariant. In particular, we show that the holographic trace anomaly in four dimension predicts the Ricci scalar squared term that would be  incompatible with the Wess-Zumino consistency condition if it were conformal. This illustrates how the foliation preserving diffeomorphic theory of gravity could be in conflict with a theorem of the dual unitary quantum field theory.
\end{abstract}

\thispagestyle{empty} 

\setcounter{page}{0}

\newpage

\section{Introduction}
A holographic approach to strongly coupled quantum field theories has now become a ubiquitous tool to understand the non-perturbative physics albeit its derivation from the first principle is yet to be established. One of the salient features of holography is that we identify the holographic direction as the one generated by the renormalization group flow (see e.g. \cite{Akhmedov:1998vf}\cite{Akhmedov:2002gq}\cite{Heemskerk:2010hk}\cite{Faulkner:2010jy} and references therein). This holographic renormalization group viewpoint seems at the heart of any attempt to derive the holography \cite{Akhmedov:2010mz}\cite{Lee:2010ub}\cite{Douglas:2010rc} e.g. AdS/CFT correspondence from quantum field theories.

From this viewpoint, while the scale transformation  plays a primary role, the special conformal transformation seems to only play a secondary role. In this paper, however, we would like to claim that the space-time diffeomorphism is crucially related to the latter. If we singled out the radial direction as a holographic renormalization group direction, the natural symmetry of the holographic theory would be a foliation preserving diffeomorphism rather than the full space-time diffeomorphism. In the Poincar\'e patch of the AdS space, for instance, the Poincar\'e isometry and the scaling isometry manifestly appear in the foliation preserving diffeomorphism while the isometry corresponding to the special conformal transformation does not preserve the Poincar\'e foliation. Therefore, naively it seems more natural to consider the holography only with a foliation preserving diffeomorphism and the resulting dual field theory without conformal invariance.

In reality, however, most of the scale invariant relativistic quantum field theories are conformal invariant \cite{Polchinski:1987dy}\cite{Dorigoni:2009ra}. Certainly most of the examples in holography (mainly from string theories) assume the conformal invariance. At the same time, some theorists believe that the  full space-time diffeomorphism is sacred in gravity. Are they somehow related?

It is still under vigorous investigations whether the scale invariance and the conformal invariance are equivalent in unitary relativistic quantum field theories. The equivalence is proved in $d=2$ \cite{Zamolodchikov:1986gt}\cite{Polchinski:1987dy} (see also \cite{Mack1}), and there exists a counterexample in $d>4$ \cite{Pons:2009nb}\cite{Jackiw:2011vz}\cite{ElShowk:2011gz}. While there are some illuminating works in $d=3, 4$ \cite{Dorigoni:2009ra}\cite{Fortin:2011ks}\cite{Fortin:2011sz}\cite{Fortin:2012ic}, the jury is still out.  Given its possibility, it is motivating to understand how scale invariant but non-conformal field theories can (or cannot) be realized holographically.

In this paper, we clarify the importance of the full space-time diffeomorphism to obtain the conformal invariance. On the other side of the same coin, we show that the holographic models based on a foliation preserving diffeomorphic theory of gravity naturally leads to a dual description of scale invariant but non-conformal field theories, assuming the existence of such.

Foliation preserving diffeomorphic theories of gravity \cite{Horava:2008ih}\cite{Horava:2009uw} have attracted their own interest from a completely different motivation from ours. They are studied as cosmological models  (see e.g. \cite{Mukohyama:2010xz} for a review).
Horava selected time as a special direction, and he proposed his foliation preserving diffeomorphic theory of gravity as a way to construct a power-counting renormalizable theory of gravity. Our model can be seen as a space-time flipped version of his gravity (with $z=1$) while the power-counting renormalizability is not necessarily assumed. We hope that our results will shed some light on possible holographic understandings of the Horava gravity\footnote{Indeed, one of his motivations to construct the foliation preserving diffeomorphic theory of gravity is to realize space-time anisotropy in holography that has potential applications in condensed matter physics while this viewpoint is rarely emphasized in the cosmological applications of his models. A recent attempt to derive the foliation preserving diffeomorphic theory of gravity as a non-relativistic (holographic) trace anomaly can be found in \cite{Griffin:2011xs}.}
 while there is always a technical as well as conceptual subtly in flipping space and time in holography.

As a concrete computation, we study the holographic trace anomaly \cite{Henningson:1998gx} of a foliation preserving diffeomorphic theory of gravity. In $d=4$,  simple dimensional analysis leads to the following form of the trace anomaly of the energy-momentum tensor (see e.g. \cite{Duff:1993wm} for a review):
\begin{align}
T^{\mu}_{\ \mu} =  a(\mathrm{Euler}) - c(\mathrm{Weyl}^2) + b R^2 + b' 
 \Box R + e \epsilon^{\rho\sigma \alpha \beta} R_{\rho\sigma\mu\nu} R^{\mu\nu}_{ \ \ \alpha \beta} + \mathrm{non \ anomalous \ terms} \label{ta}
\end{align}
In almost all literatures on the trace anomaly, only $a$ and $c$ are studied: in particular, the central charge $a$ plays an important role in understanding the  so-called $a$-theorem that states the existence of the monotonically decreasing function along the renormalization group flow \cite{Cardy:1988cwa}\cite{Jack:1990eb}\cite{Komargodski:2011vj}\cite{Nakayama:2011wq}\cite{Komargodski:2011xv}. The geometric meaning of these terms are clarified in \cite{Deser:1993yx}.
On the other hand, $e$ vanishes for any CP preserving theories \cite{Deser:1980kc}\cite{Deser:1996na},\footnote{Its significance was revisited in \cite{Nakayama:2012gu}.} and $b'$ can be removed by adding the local counterterm $\int d^4x \sqrt{g}R^2$ in the action so they may be discarded for our purposes. Finally, $bR^2$ term does not satisfy the Wess-Zumino consistency condition \cite{Bonora:1983ff}\cite{Bonora:1985cq}\cite{Cappelli:1988vw}\cite{Osborn:1991gm}, so it cannot appear in the trace anomaly if the theory is conformal (i.e. under the absence of the last non-anomalous terms in \eqref{ta} up to a possible improvement).

We, however, show that the holographic trace anomaly of a foliation preserving diffeomorphic theory of gravity predicts non-zero $b$, which means that the theory must be scale invariant but not conformally invariant: otherwise it is inconsistent. Indeed, as we have mentioned, the theory does not admit the foliation non-preserving diffeomorphism that would generate the special conformal transformation. 

Given the recent interest in finding a scale invariant but not conformal invariant field theory in $d=3,4$, we believe our holographic construction will serve as a good exercise to uncover its peculiar property and existence itself. 
Also given the fact that scale but non-conformal field theory does exist in $d>4$, it is urgent how it can fit with the holography, and our model provides a certain prototype.
 Finally, the proof of the non-existence of scale invariant but not conformal invariant field theories in $d=2$ may suggest that the foliation preserving diffeomorphic theory of gravity could be in conflict with the idea of the holography. In each cases, the study of the holographic models seems useful.

The organization of the paper is as follows. In section 2, we review a free $U(1)$ Maxwell field theory in $d>4$ as a simple example of scale invariant but not  conformal invariant field theories, and we try to construct its holographic dual. We see how the foliation preserving diffeomorphism is natural while the full space-time diffeomorphism is not. In section 3, we study the holographic renormalization group of a foliation preserving diffeomorphic theory of gravity. We compute the holographic trace anomaly and find the non-zero Ricci squared term in $d=4$. In section 4, we present some discussions for future works.

\section{Holographic dual of free Maxwell theory?}
As a canonical example of  scale invariant but not conformal invariant field theories, we will study a free $U(1)$ Maxwell field theory in higher dimension $d>4$  \cite{Jackiw:2011vz}\cite{ElShowk:2011gz}. The theory is unitary and scale invariant, but the correlation functions among gauge invariant field strengths do not satisfy the Ward-Takahashi identity for the conformal invariance by treating field strength as a conformal primary operator. The energy-momentum tensor of the free $U(1)$ Maxwell action ($F_{\mu\nu} = \partial_\mu A_\nu - \partial_\nu A_\mu$ as usual)
\begin{align}
S = \int d^d x \frac{1}{4} F^{\mu\nu} F_{\mu\nu} 
\end{align}
has the non-trivial divergence of the Virial current in its trace (in $d>4$)
\begin{align}
T^{\mu}_{\ \mu} = \frac{4-d}{4}F^{\mu\nu} F_{\mu\nu} = \frac{4-d}{2} \partial^\mu (A^{\nu}F_{\nu\mu}) \ ,
\end{align}
where $J_\mu = \frac{4-d}{2} A^\nu F_{\nu\mu}$ is the Virial current. The existence of the non-trivial Virial current states that the theory is not conformal.
We may be able to embed the theory in a conformal field theory by introducing ``vector potential" in a suitable gauge, but the embedding is non-unitary \cite{ElShowk:2011gz}.

In \cite{Douglas:2010rc}, a recipe to construct a holographic dual of a free scalar field theory was presented. The construction heavily relies on the conviction that the renormalization group scale will be given by the  ``radial direction" under holography, which is the basis of the holographic renormalization. From this perspective, the notion of the conformal invariance, or more precisely symmetry under the special conformal transformation only plays a secondary role. Nevertheless, at the end, the vacuum solution of their equations of motion has the full AdS isometry, and the theory is invariant under the special conformal transformation, which is expected because the free massless scalar is conformally invariant.

In this section, we repeat a similar holographic construction of the holographic dual theory from the scale invariant but not conformal invariant free $U(1)$ Maxwell field theory (in $d>4$).  There is no essential technical difference from the scalar case \cite{Douglas:2010rc}, but it is instructive to see how the ``emergence" of the AdS space-time isometry does not occur in our example.

The starting action is the momentum space free $U(1)$ Maxwell field theory given by
\begin{align}
\int d^dp d^dq\left\{P_{\mu\nu}(p,q)-B_{\mu\nu}(p,q)\right\}A^{\mu}(p) A^{\nu}(q) \ ,
\end{align} 
where $P_{\mu\nu}$ is the cut-off kinetic term with suitable gauge fixing. The details of the gauge fixing would not affect the physics, and for simplicity, we take the Feynman gauge\footnote{Actually, the following argument would not depend on the fact that the theory is unitary, so we may allow an arbitrary kinetic term for the vector field. See e.g. \cite{Riva:2005gd}\cite{Nakayama:2010ye} how the generic choices of the kinetic term break conformal invariance while preserving the scale invariance.} 
\begin{align}
P_{\mu\nu} = \frac{p^2}{K(p^2/\Lambda^2)} \delta^{d}(p-q) \eta_{\mu\nu} \ .
\end{align}
Here the cut-off function $K$ vanishes for large momenta and becomes constant for small momenta.
The exact renormalization group equation determines the flow of the generalized coupling $B_{\mu\nu}$ as
\begin{align}
d_\Lambda B_{\mu\nu}(p,q) = - \int d^ds \frac{1}{s^2} \frac{\partial K(s^2/\Lambda^2)}{\partial \Lambda}B_{\mu\rho}(p,s)B^{\rho}_{\ \nu}(s,q) \ . \label{rge}
\end{align}
We will identify the renormalization group scale $\Lambda$ with the holographic direction: $\Lambda  =\frac{1}{r}$.

We can immediately see that the construction here is identical to that of \cite{Douglas:2010rc} except that the field $B$ is now a matrix with Lorentz indices. We expand the field $B$ and the cut-off differential propagator $\alpha^{\mu\nu} = \frac{d_\Lambda K(p^2/\Lambda^2) \eta^{\mu\nu}}{p^2} \delta^d(p-q)$ with respect to momenta $p$ and $q$ as
\begin{align}
B^{\mu\nu}(p/\Lambda, q/\Lambda) &= \sum_{s,t=0}^\infty \Lambda^{-s-t} B^{\mu\nu}_{a_1 \cdots a_s, b_1 \cdots b_t}p^{a_1} \cdots p^{a_s} q^{b_1}\cdots q^{b_t} \cr
&\equiv \Lambda^{-s-t}B_{\underline{s}\underline{t}}^{\mu\nu}p^{\underline{s}}q^{\underline{t}} \cr
\alpha_{\underline{s}\underline{t}}^{\mu\nu} &= \Lambda^{2-d-s-t} \int d^dp d^d q \alpha^{\mu\nu}(p,q)p_{\underline{s}}q_{\underline{t}} \ .
\end{align}
Here, we are following the notation used in \cite{Douglas:2010rc}.

By using these components, one can introduce the connections $W$ and $\tilde{W}$ that take values in the ``non-commutative" space.
\begin{align}
W_M^{\mu\nu} &= \frac{1}{r}P_M\eta^{\mu\nu} + B_{\underline{q}\underline{s}}^{\mu\rho}[\alpha^{\underline{s}\underline{p}}_M]_{\rho}^{\ \nu} \cr
\tilde{W}_M^{\mu\nu} &= \frac{1}{r} P_M\eta^{\mu\nu} \ . \label{noncc}
\end{align}
Here $P_r$ generates the dilatation, and $P_\mu$ generates the translation so that the connection $\frac{1}{r} P_M$ satisfies the AdS algebra (see \cite{Douglas:2010rc} for details) under $*$ product. We also introduced $[\alpha^{\underline{s}\underline{p}}_{M=r}]_{\rho}^{\ \nu} = [\alpha^{\underline{s}\underline{p}}]_{\rho}^{\ \nu}$, and $[\alpha^{\underline{s}\underline{p}}_{M=\mu}]_{\rho}^{\ \nu} = 0$. Consult \cite{Vasiliev:1999ba}\cite{Douglas:2010rc} for how to relate the non-commutative fields and their component with $\underbar{s}\underbar{q}$ indices assumed in \eqref{noncc}.

The renormalization group equation \eqref{rge} can be obtained as a particular solution of
\begin{align}
\partial_M B_{\mu\nu} + W_{M \mu\rho}* B^{\rho}_{\ \nu} - B_{\mu\rho}* \tilde{W}_{M \ \nu}^{ \ \ \rho} = 0\ 
\end{align}
with \eqref{noncc}. Here, the connections are both flat.
\begin{align}
dW_{\mu\nu} + W_{\mu\rho} \wedge * W^{\rho}_{\ \nu} = 0 \ , \ \ d\tilde{W}_{\mu\nu} + \tilde{W}_{\mu\rho} \wedge * \tilde{W}^{\rho}_{\ \nu} = 0 \ .
\end{align}

The details of the construction, which we have omitted, are not important in our discussions, but let us emphasize the salient features: (1) the construction can be done in any scale invariant but not conformal invariant free field theories, (2) since the connection as well as $B$ field has $d$-dimensional Lorentz indices (i,e, $\mu, \nu$) rather than $d+1$ dimensional indices (i.e. $M, N$), the manifest $d+1$ dimensional diffeomorphism (or local Lorentz transformation) is broken. This is in stark contrast with the scalar case, where the $d+1$ dimensional local Lorentz invariance emerges in the final equations of motion albeit the physical origin of the requirement to choose the particular solution that gives the free scalar field correlation functions is unclear. For instance, even in the scalar case, the analogue of the decomposition \eqref{noncc} breaks the AdS invariance.  The particular choice also breaks the full space-time diffeomorphism down to the foliation preserving diffeomorphism.

In the construction of \cite{Douglas:2010rc}, the emergence of the full space-time diffeomorphism, or the local Lorentz invariance is quite artificial. Since the renormalization group flow direction $r$ has its very different origin than the flat Minkowski plane spanned by $x^\mu$ foliating the space-time, it is far from obvious why these coordinates must behave in a unified way under the full space-time diffeomorphism. A more natural symmetry would be the foliation preserving diffeomorphism. Indeed, in the free $U(1)$ Maxwell field theory case, the only manifest symmetry of the vacuum is the foliation preserving diffeomorphism. We will discuss how the emergence (or the requirement) of the full space-time diffeomorphism is crucially related to the conformal invariance of the dual theory.

Before moving on, let us make a comment on the lower dimensional case, when the free $U(1)$ Maxwell field theory does possess the conformal invariance.
In $d=3$, a free $U(1)$ gauge field is dual to a free scalar. If we had utilized the free field equivalence: $F_{\mu\nu} = \epsilon_{\mu\nu\rho} \partial^\rho \phi$, and rewritten the renormalization group equation with respect to a free scalar field $\phi$ and its scalar source $B$ instead of $A_\mu$ with the gauge invariant source term, we would have ended up with the AdS vacuum solution with the full  space-time diffeomorphism. This is in accord with the fact that the free $U(1)$ Maxwell field theory in $d=3$ can be unitarily embedded in a conformal field theory by the duality transformation. The effect of the duality is non-trivial in the holographic construction we have presented.

The situation in $d=4$ is less clear. Although the constructed holographic theory does possess the conformal invariance, the above construction does not automatically lead to the manifest AdS isometry and the space-time diffeomorphism. Unlike in the $d=3$ case, the duality would not help either since the theory is self-dual. We need a better understanding of the emergence of the AdS isometry here because, after all, the free field construction of the $\mathcal{N}=4$ super Yang-Milles theory must be available and the result should be fully AdS invariant. Presumably, the  $\mathcal{N}=4$ supersymmetric formulation will solve the problem naturally because we may regard the free Maxwell field as a descendant of the scalar component under the $\mathcal{N}=4$ supersymmetry.\footnote{However, we still have to understand how this mechanism would not work in $d>4$.} Our main interest in this paper is not the recovery of the conformal invariance, but rather how it is broken, so we will leave this question for future investigations.

\section{Holographic renormalization of foliation preserving gravity}
As we have discussed, the holographic renormalization group flow singles out the radial direction as a distinguished coordinate. In any attempt to derive the holographic description of the field theories, this seems unavoidable. We have studied one particular attempt to construct a holographic dual theory from the free $U(1)$ Maxwell field theory in the last section.
Since we have singled out the particular direction, the role played by the full space-time diffeomorphism, let alone its origin, is less clear. Indeed, in the example studied in the last section, the existence of the full space-time diffeomorphism is doubtful (when $d>4$) unlike the scalar case.
In other words, it is logically possible that the holographic theory is only invariant under the foliation preserving diffeomorphism but is not invariant under the full space-time diffeomorphism.

We, however, claim that the full space-time diffeomorphism is  crucial in order to understand the special conformal invariance of the dual field theory at the fixed points. Alternatively, abandoning the full space-time diffeomorphism is equivalent to abandoning the conformal symmetry. This can be heuristically understood if we look at the space-time metric of the AdS space in the Poincar\'e patch:
\begin{align} 
ds^2 &= l^2\frac{dr^2 + \eta_{\mu\nu} dx^\mu dx^\mu}{r^2} \ .
\end{align}
The metric has the manifest Poincar\'e isometry together with the scaling isometry\begin{align}
r\to \lambda r \ , \ \ x^\mu \to \lambda x^\mu \ .
\end{align}
In addition, the metric is invariant under the isometry corresponding to the special conformal transformation with an infinitesimal parameter $k^\mu$:
\begin{align}
\delta_{k} r = 2(\eta_{\rho\sigma}k^\rho x^\sigma)r \ , \ \ \delta_{k} x_\mu = 2(\eta_{\rho\sigma}k^\rho x^\sigma)x_\mu - (r^2 + \eta_{\rho\sigma}x^\rho x^\sigma) k_\mu \ .  \label{spc}
\end{align}
A crucial observation here is that this somewhat obscured symmetry of the metric
requires the non-trivial mixing of the radial direction and the foliating space-time.

To make this point clearer, we would like to study the holography of a foliation preserving diffeomorphic theory of gravity in this section. Recently, foliation preserving diffeomorphic theories of gravity have been studied  a lot in the context of a possible power-counting renormalizable theory of quantum gravity, which was initiated by Horava in his seminal work \cite{Horava:2008ih}\cite{Horava:2009uw}. Our motivation is quite different, but we may regard our holographic model as a space-time flipped version of the Horava gravity.

The basic idea is to write the $d+1$ dimensional space-time metric in the ADM(-like) form as
\begin{align}
ds^2 = N^2 dr^2 + G_{\mu\nu}(dx^\mu + N^\mu dr)(dx^\nu + N^\nu dr) \ .
\end{align}
We singled out the radial direction $r$ (instead of ``time" in the usual ADM construction), which is supposed to be dual to the renormalization group scale. If we are interested in the holographic dual of a Lorentzian signature quantum field theory, the $d$-dimensional metric $G_{\mu\nu}$ must possess the Lorentzian signature. 
The gravitational theory that describes the dynamics of the metric must be invariant under the infinitesimal foliation preserving diffeomorphism
\begin{align}
\delta r = f(r) \ , \ \ \delta x^\mu = \xi^\mu(r,x^\mu) \ ,
\end{align}
under which
\begin{align}
\delta N &= \partial_r (Nf) \cr
\delta N^\mu &= \partial_r (N^\mu f) + \partial_r \xi^\mu + \mathcal{L}_\xi N^\mu \cr
\delta G_{\mu\nu} &= f \partial_r G_{\mu\nu} + \mathcal{L}_\xi G_{\mu\nu} \  ,
\end{align}
where $\mathcal{L}_\xi$ is the $d$-dimensional Lie derivative with respect to the vector $\xi$.

An important observation here is that although the scale transformation and the Poincar\'e transformation are foliation preserving diffeomorphisms, the special conformal transformation of the AdS space \eqref{spc} is not. Thus, the holographic dual theories of a foliation preserving diffeomorphic theory of gravity must be scale invariant and Poincar\'e invariant but not conformal invariant.

As an illustration, we will consider the following action for a particular foliation preserving  diffeomorphic theory of gravity:
\begin{align}
 S = \int N dr \sqrt{-G} d^d x (K^{\mu\nu} K_{\mu\nu} - \lambda K^2 + R + \Lambda) \ .
\end{align}
Here $K_{\mu\nu} = \frac{1}{2N} (\partial_r G_{\mu\nu} - D_\mu N_\nu - D_\nu N_\mu)$, and $K = G^{\mu\nu} K_{\mu\nu}$. $D_\mu$ is the covariant derivative with respect to the $d$-dimensional metric $G_{\mu\nu}$ and $R$ is the $d$-dimensional Ricci scalar. The main goal of this section is to study the holographic renormalization group flow of this foliation preserving diffeomorphic theory of gravity to uncover the scale invariant but non-conformal nature of the dual field theory.

The free dimensionless parameter $\lambda \neq 1$ here breaks the space-time diffeomorphism to the foliation preserving diffeomorphism. When $\lambda = 1$, the action formally recovers the full space-time diffeomorphism and it coincides with the Einstein-Hilbert action with a cosmological constant (up to surface terms).

The action is the space-time flipped version of the ``IR-limit of the Horava gravity" (with IR scaling $z=1$). In the Horava gravity, we typically introduce  higher derivative ``potential terms" (e.g. $R^n$ with $n\ge2$) to improve the renormalizability of the theory. Here, we note that such terms might lead to a non-unitary spectrum because $R$ in our space-time flipped version contains the time derivative  unlike in the Horava gravity where the foliating space has the Euclidean signature.
We restrict ourselves to the case $z=1$, but if we dare to abandon the unitarity or if we work on the Euclidean gravity, there is nothing wrong with the inclusion of the higher derivative terms at least at the classical level.\footnote{The inclusion of the higher derivative terms will naturally lead to the violation of $a=c$ of the four-dimensional holographic central charge. The effect of higher derivative terms in the holographic trace anomaly of fully space-time diffeomorphic theories can be found in \cite{Nojiri:1999mh}\cite{Blau:1999vz}.}

The equation of motion becomes
\begin{align}
 0 = &-\frac{1}{N} (\partial_r - N^\sigma D_\sigma) p_{\mu\nu} + \frac{1}{N}(p_{\mu \sigma} D_\nu N^\sigma + p_{\nu \sigma} D_\mu N^{\sigma}) \cr 
 &- K p_{\mu\nu} + 2K_{\mu}^{\sigma} p_{\sigma \nu} + \frac{1}{2}\Lambda g_{\mu\nu} - \left(R_{\mu\nu} -\frac{R}{2}G_{\mu\nu}\right) \ ,
\end{align}
where  $p_{\mu\nu} = K_{\mu\nu} - \lambda K g_{\mu\nu}$ is the ``canonical momentum". We augment it with the ``momentum constraint"
\begin{align}
0 = D^\mu p_{\mu\nu} \ , 
\end{align}
and the ``Hamiltonian constraint"\footnote{This equation is obtained by assuming the ``Lapse function" $N$ depends on both $r$ and $x^\mu$. If we introduced the projectability condition that demands the ``Lapse function" only depends on the radial direction $r$, then the Hamiltonian constraint \eqref{Ham} will be replaced by the integrated form $0 = \int d^dx \sqrt{-g} \left(K^{\mu\nu} p_{\mu\nu} - \Lambda - R\right)$. We may regard the local form \eqref{Ham} as an ansatz to solve the metric equation in the projectable case.}
\begin{align}
0 = K^{\mu\nu} p_{\mu\nu} - \Lambda - R \label{Ham}
\end{align}

Now we would like to understand the holographic renormalization group flow of the foliation preserving diffeomorphic theory of gravity. In particular, we would like to compute the holographic trace anomaly to see the effect of the scale invariance without conformal invariance. 
The argument here follows the standard recipe to compute the holographic trace anomaly in the full space-time diffeomorphic theory of gravity \cite{Henningson:1998gx}.
We first introduce the Graham-Fefferman ansatz (using the coordinate $\rho = r^2$)\footnote{Since there is no theorem that guarantees that we can take this coordinate in foliation preserving diffeomorphic theories of gravity unlike in the fully space-time diffeomorphic theories of gravity, we emphasize that this is merely an ansatz.}
\begin{align}
ds^2 = l^2\left(\frac{d\rho^2}{4\rho^2} + \frac{g_{\mu\nu}(\rho,x) dx^{\mu}dx^{\nu}}{\rho}\right) 
\end{align}
with the power series expansion for the metric near the boundary $\rho = 0$:
\begin{align}
g = g^{(0)} + \rho g^{(2)} + \cdots + \rho^{d/2} g^{(d)} + \rho^{d/2} \log \rho h^{(d)} + \mathcal{O}(\rho^{d/2+1}) \ . \label{ansatz}
\end{align}
In this paper, we are only interested in even $d$, where the trace anomaly is non-zero, and in the above expansion it was assumed as such. Note that $g_{\mu\nu} = \eta_{\mu\nu}$, which gives the AdS space in Poincar\'e patch, is a vacuum solution when $\Lambda = d(1-\lambda d)/l^2$.

With the Graham-Fefferman ansatz, the equations of motion of the foliation preserving diffeomorphic theory of gravity become
\begin{align} 
\rho\left(-2g''+2\lambda\mathrm{Tr}(g^{-1}g'' )g + (1-2\lambda)\mathrm{Tr}(g^{-1}g' g^{-1}g') g 
+ 2g'g^{-1}g' - \mathrm{Tr}( g^{-1}g')g'\right) + &\cr
+ (d-2)g' + (2\lambda-1)\mathrm{Tr}(g^{-1} g') g - \mathrm{Ric}[g]  &= 0  \cr
(g^{-1})^{\mu\sigma}(\lambda D_\nu g'_{\mu \sigma} - D_{\sigma} g'_{\mu\nu}) &= 0  \cr
(2d\lambda-2)\mathrm{Tr}(g^{-1}g'') + (d-2\lambda d + 1)\mathrm{Tr}(g^{-1}g'g^{-1}g') + (-1+ \lambda) (\mathrm{Tr}(g^{-1}g'))^2 &= 0 \ . \cr \label{eomm}
\end{align}
Here we have used the relation $\Lambda = d(1-\lambda d)/l^2$ between the cosmological constant and the AdS radius.

We may substitute the ansatz \eqref{ansatz} into \eqref{eomm} and solve the metric perturbatively with respect to $\rho$ with a given boundary metric $g^{(0)}$ up to $\mathcal{O}(\rho^{d/2})$.
The first order solution, for instance, becomes
\begin{align}
g^{(2)}_{\mu\nu} = \frac{1}{d-2} \left(R_{\mu\nu}^{(0)} - \frac{(2\lambda-1)R^{(0)}}{2d\lambda- 2} g_{\mu\nu}^{(0)} \right) \ . \label{solu}
\end{align}
Here $R^{(0)}$ is the curvature tensor constructed from the metric $g^{(0)}$.

The holographic trace anomaly can be computed by employing the same technique introduced in \cite{Henningson:1998gx} when the theory has the full space-time diffeomorphism. For this purpose, we need to know the logarithmic divergence of the on-shell action with the cut-off at $\rho = \epsilon$. As in the Einstein case, the logarithmic divergence only comes from the bulk integral 
\begin{align}
S&= l^{d/2+1}\int_{\epsilon} d\rho d^{d}x \frac{\rho^{-d/2-1}}{2} \sqrt{-g} \left[2(\Lambda +R)\right] |_{\log\epsilon} \ .
\end{align}
We expand the metric $g$ and curvature $R$ with \eqref{ansatz} and the explicit solutions such as \eqref{solu} to obtain the $\rho^{2/d}$ term inside the on-shell action integral. In the $d=4$ case we are most interested in, the result is 
\begin{align}
S & = \frac{l^3}{4}\log \epsilon  \int d^4 x \sqrt{-g_0} \left(R_{\mu\nu}^{(0)}R^{\mu\nu (0)} - \frac{\lambda}{4\lambda-1}R^{(0)2}\right)  \ .
\end{align}

The logarithmic dependence on $\epsilon$ gives rise to the holographic trace anomaly of the dual theory.
The result in $d=4$ is
\begin{align}
\left\langle T^{\mu}_{\ \mu} \right\rangle &= -2c \left(R_{\mu\nu} R^{\mu\nu} - \frac{\lambda}{4\lambda -1} R^2 \right) \cr
& = c \left((\mathrm{Euler} - \mathrm{Weyl}^2) -\frac{2}{3}\frac{\lambda-1}{4\lambda-1}R^2\right) \ .
\end{align}
Here $\mathrm{Euler} = R^{\mu\nu\rho\sigma} R_{\mu\nu\rho\sigma} - 4R^{\mu\nu} R_{\mu\nu}+ R^2$ is the Euler density, and $\mathrm{Weyl}^2=R^{\mu\nu\rho\sigma} R_{\mu\nu\rho\sigma} - 2R^{\mu\nu} R_{\mu\nu}+ \frac{1}{3}R^2$ is the Weyl tensor squared. $c \propto l^3$ is the holographic central charge of the dual field theory whose value is the same as when $\lambda = 1$ as long as the AdS radius $l$ is fixed.

As in the Einstein case, the holographic trace anomaly with the two derivative action here predicts $a=c$ irrespective of the value of $\lambda$. The additional term with the Ricci scalar squared is something new. Actually, it is known that this $R^2$ term  does not satisfy the Wess-Zumino consistency condition for the Weyl anomaly  \cite{Bonora:1983ff}\cite{Bonora:1985cq}\cite{Cappelli:1988vw}\cite{Osborn:1991gm}, so it is not allowed for conformal field theories. A simplified explanation is that since the Weyl transformation is Abelian, the Weyl anomaly of the Weyl anomaly (up on integration) must vanish. This is the case for the Euler density as well as all the Weyl invariants such as the Weyl tensor squared or Hirzebruch-Pontryagin density, but it is not true for the $R^2$ term.
Of course, the Wess-Zumino consistency condition is obtained by demanding the flat space conformal invariance and it is presumably violated here, so there is no immediate theoretical inconsistency. It merely dictates that the dual field theory cannot be conformal invariant when $\lambda \neq 1$. Note that this $R^2$ term vanishes at $\lambda =1$ as expected from the Einstein gravity, which is supposed to be dual to a conformal field theory.

We can repeat the same analysis in other dimensions. In $d=2$, the holographic trace anomaly is still given by the Euler density: $\langle T^{\mu}_{\ \mu} \rangle = -\frac{c}{12}R$, and the effect of $\lambda$ is only changes the value of the central charge through the relation between $\Lambda$ and the AdS radius $l$. Although we cannot see the violation of the conformal invariance from the trace anomaly, as is also the case with the field theory analysis, we have no evidence for the conformal invariance of the dual field theory here when $\lambda \neq 1$ because the isometry corresponding to the special conformal invariance is lost.

For instance, we may try the  Brown-Henneaux  approach \cite{Brown:1986nw} to construct the asymptotic symmetry generators corresponding to the Virasoro symmetry. The first difficulty is that we cannot freely switch the Poincar\'e coordinate to the global coordinate. The coordinate transformation between the Poincar\'e coordinate and the global coordinate does not belong to the foliation preserving diffeomorphism. We can easily check that the AdS space with the global coordinate (i.e. $ds^2 = l^2\left(\frac{d\rho^2}{4\rho^2} + \frac{(1-\rho)^2 d\phi^2 - (1+\rho)^2 dt^2}{\rho} \right)$ ) does not solve the equations of motion when $\lambda \neq 1$ while the Poincar\'e patch does solve them.

Furthermore, most of the asymptotic symmetry generators corresponding to would-be Virasoro symmetry do not preserve the foliation, so it is impossible to construct the Virasoro generators
\begin{align}
\zeta_n^{\pm} = e^{in(t\pm \phi)}\left(\partial_{\pm} -\frac{n^2}{2r^2} \partial_{\mp} - \frac{in r}{2}\partial_r \right) \ .
\end{align}
 In particular, there is no corresponding foliation preserving asymptotic symmetry generator corresponding to the special conformal transformation, which would be $n=-1$, in this approach (even if we restrict ourselves to the Poincar\'e coordinate).

In higher dimensions, say $d=6, 8 \cdots$, although we have not performed the detailed computation, it is straightforward to repeat our analysis and compute the holographic trace anomaly. It is expected that the trace anomaly contains the terms that are not allowed by the Wess-Zumino consistency condition for conformal field theories. We have many candidates of these terms such as $R^{d/2}$. 
 Again, this will not lead to an immediate inconsistency because the theory is expected to be only scale invariant but not conformal invariant. Indeed, it seems very likely that the free $U(1)$ Maxwell field theory in $d=6,8\cdots$ would produce the trace anomaly which is not allowed by the Wess-Zumino consistency condition.

However, we know that in $d=2$, the extra constraint from the unitarity and the discreteness of the spectrum demands that scale invariant field theory must be conformal invariant \cite{Polchinski:1987dy}. This suggests that holographic dual of the foliation preserving diffeomorphic theories of gravity must be non-unitary, if any, in $d=2$. In $d=4$, we have no examples of scale invariant but not conformal invariant field theory yet, so the holographic theory dual might be pathological as in $d=2$. In $d>4$,  uniatry scale invariant but not conformal invariant field theories do exist as we have seen in section 2, so the holographic description depicted there may well possess no full space-time diffeomorphism but only have a foliation preserving diffeomorphism.

\section{Discussions}

In this paper, we have studied holography of a foliation preserving diffeomorphic theory of gravity. This is motivated by the conviction that any derivation of the holography singles out the radial direction as the renormalization scale, so there is no a-priori reason why the full space-time diffeomorphism must be imposed. As we have seen in section 2, it seems more natural to assume a foliation preserving diffeomorphism.
The consequence is that the dual theory, if any, is only scale invariant but not conformal invariant. 

Our result can be regarded as the first demonstration of the possibility to generate the Ricci scalar squared term in the holographic trace anomaly at the fixed points. We emphasize again that the Ricci scalar squared term in the trace anomaly is forbidden by the Wess-Zumino consistency condition in conformal field theories. There is no theoretical inconsistency here because our dual field theory must be only scale invariant but not conformal invariant. 

We would like to mention some unusual features of the $R^2$ term in the holographic trace anomaly. First of all, this already appears at the second derivative theories of  gravity in their classical limit. Thus, the effect even survives in the large $N$ limit of the  gauge/gravity correspondence. This should be contrasted with the deviation from $c=a$ due to the higher derivative corrections. Secondly,  since it does not satisfy the Wess-Zumino consistency condition, the term cannot be integrated to generate the Wess-Zumino term even if it is spontaneously broken unlike $a$ or $c$ anomalies \cite{Komargodski:2011vj}.

The explicit breaking of the space-time diffeomorphism down to the foliation preserving diffeomorphism is related to the spontaneous breaking of the Lorentz symmetry (or more precisely AdS symmetry), and our treatment in this paper will be related to the models studied in  \cite{Nakayama:2009qu}\cite{Nakayama:2009fe}\cite{Nakayama:2010wx}\cite{Nakayama:2011zw}, where the  holographic theories for scale invariant but not conformal invariant field theory was initiated. It would be very interesting to compute the holographic trace anomaly there and compare it with the results presented in this paper to understand the relation between the foliation preserving diffeomorphic theory of gravity and the spontaneous Lorentz symmetry (AdS symmetry) breaking.

With this regards, in \cite{Nakayama:2012gu}, a potential CP odd term in the trace anomaly, given by the Hirzebruch-Pontryagin density in $d=4$, was pursued. In particular, the holographic model that would generate such a term was constructed with the help of the spontaneous AdS symmetry breaking. It seems easier to introduce such a CP odd term in the holographic trace anomaly if we consider the foliation preserving diffeomorphic theory of gravity rather than in the Einstein gravity with the AdS symmetry breaking.\footnote{The latter necessitates the violation of the (strict) null energy condition, so apparently the foliation preserving diffeomorphic theory of gravity might be pathological in some ways if we assume the equivalence. It seems important to understand precisely in which aspect the theory is pathological. This ultimately leads to the question whether the scale invariant but not conformal invariant field theory is theoretically satisfactory or not.} The lack of the full space-time diffeomorphism allows us to introduce the four-dimensional Levi-Civita symbol (rather than five-dimensional one) in the action. For instance, the term like $\int Ndr \sqrt{-G}d^4x  K \epsilon^{\rho\sigma \alpha \beta} R_{\rho\sigma\mu\nu}R^{\mu\nu}_{\ \ \alpha\beta}$ would naturally give rise to the holographic trace anomaly proportional to the Hirzebruch-Pontryagin density.

While we have concentrated on the holographic trace anomaly in this paper, there are other modifications that we expect in the holography of foliation preserving diffeomorphic theories of gravity compared with the conventional full space-time diffeomorphic theories of gravity. We would like to close this paper by mentioning some of them in relation to their field theory implications.

First of all, in $d=2$, the scale invariant but not conformal invariant field theory must be non-unitary (with other technical assumptions). Therefore,  holographic dual of foliation preserving diffeomorphic theories of gravity may be pathological at least in $d=2$. Since we have proposed an explicit gravity model, it seems very interesting to compute the energy-momentum tensor correlation functions from holography and compare them with the field theory proof of the equivalence between scale invariance and conformal invariance in $1+2$ dimension.\footnote{We can repeat a completely parallel discussion for a chiral version of the equivalence between scale invariance and conformal invariance \cite{Hofman:2011zj}. The gravitational theorem again assumes the full diffeomorphism in \cite{Nakayama:2011fe}. The symmetry enhancement would not occur in foliation preserving diffeomorphic theories of gravity.}

Secondly we would like to revisit the holographic $c$-theorem \cite{Girardello:1998pd}\cite{Freedman:1999gp}. The holographic $c$-theorem predicts that there exists a monotonically decreasing function of $r$ along the holographic renormalization group flow. The derivation is based on the null energy-condition together with the absence of the ghost in space-time diffeomorphic theory of gravity \cite{Myers:2010xs}\cite{arXiv:1011.5819}. In foliation preserving diffeomorphic theories of gravity, the notion of the null energy condition is less clear. In particular, we typically need to evaluate $T^{r}_{\ r} - T^{t}_{\ t}$ to derive the holographic $c$-theorem, but since $r$ direction is distinguished in the foliation preserving diffeomorphic theories of gravity, it is not so obvious why this must be constrained. Coincidentally, we have observed a possible failure of the proof of the $a$-theorem in $d=4$ when the fixed points are scale invariant but not conformal invariant \cite{Nakayama:2011wq}\cite{Komargodski:2011xv}. It is worth investigating this point further.

Another interesting area of study is the entanglement entropy. At the conformal fixed point, it is known that the universal part of the entanglement entropy is related to the trace anomaly. In particular, in $d=4$, they are characterized by $a$ and $c$ depending on the entanglement surface \cite{Ryu:2006ef}\cite{Solodukhin:2008dh}\cite{Casini:2011kv}. However, the derivation relies on the conformal invariance, and it is less known what happens when the theory is only scale invariant but not conformal invariant.\footnote{Also, the CP-odd Hirzebruch-Pontryagin density, which is compatible with the conformal invariance, is almost always neglected.} Do we expect a non-trivial dependence on the $R^2$ term in the trace anomaly? Or, is there any hidden constraint from the entanglement entropy that forbids the $R^2$ term in the trace anomaly (see e.g. \cite{Schwimmer:2008yh}\cite{Banerjee:2011mg} for the Wess-Zumino consistency condition of the holographic renormalization and entanglement entropy)? In order to address these questions, it seems urgent to reconsider the holographic approach to the entanglement entropy \cite{Ryu:2006bv}\cite{Ryu:2006ef}\cite{Casini:2011kv} in foliation preserving diffeomorphic theories of gravity. All these would lead to better understandings of scale invariant but not conformal invariant field theories and deep aspects of holography and quantum gravity.

\section*{Acknowledgements}
The author would like to thank S.~Mukohyama for insightful discussions on the Horava gravity. He also would like to acknowledge N.~Ogawa for discussions.
The work is supported by the World Premier International Research Center Initiative of MEXT of Japan. 

\appendix

\section{Holographic Virial current and some comments on recent literatures}

In scale invariant but non-conformal field theories, the non-trivial Virial current plays a significant role. In this appendix, we would like to study its realization from holography. We also make some comments on the connection between (holographic) c-theorem and (non-)existence of scale invariant but non-conformal field theories in $(1+3)$ dimension.\footnote{The author would like to thank H.~Osborn for the discussions presented in this appendix.}

To understand the relation between the vector condensation in holography and the non-trivial existence of the Virial current, we begin with the holographic renormalization group\footnote{Non-trivial vector condensation in our discussion effectively reduces to the foliation preserving diffeomorphic theory of gravity in the bulk.}:
\begin{align}
ds^2 = e^{2A(r)} dx^\mu dx_\mu + dr^2 \ ,
\end{align}
in conventional Einstein gravity.
The holographic central charge $a = \pi^{d/2}/\Gamma(d/2)(A')^{d-1}$ satisfies
\begin{align}
\frac{da}{dr} = \frac{\pi^{d/2}}{\Gamma(d/2)(A')^d} (T^{r}_{r} - T^t_t) \  ,
\end{align}
where we used the Einstein equation.
If the theory satisfies the null energy condition, the strong c-theorem $\frac{da}{dr} \ge 0$ holds. To understand the  strongest c-theorem, let us take the non-linear sigma model for the matter action. 
The right hand side becomes
\begin{align}
T^{r}_{r} - T^{t}_t = G^{IJ} \partial_r \Phi_I \partial_r \Phi_J \ .
\end{align}
It is natural to regard $\partial_r \Phi_I$ as the beta function $\beta_I$ and $G^{IJ}$ as the Zamolodchikov metric. Thus, it agrees with the field theory expectation
\begin{align}
\frac{da}{dr} = G^{IJ} \beta_I \beta_J \ 
\end{align}
whose field theory derivation was discussed by Osborn \cite{Osborn:1991gm} in detail.
If we assume that the Zamolodchikov metric is positive definite (in compatible with the strict null energy condition \cite{Nakayama:2010wx}), we obtain the strongest c-theorem.

To look for a possibility of scale but non-conformal invariance, or the cyclic renormalization group flow, it is crucial to realize the operator identity such as $\beta_I O^I = \partial^\mu J_\mu$ in holography. The analogous identification is naturally realized by gauging the bulk scalar field (i.e. replace $\partial_\mu$ by $D_\mu = \partial_\mu + iA_\mu$). Suppose we replace the kinetic term $G^{IJ}\partial^M \Phi_I \partial_M \Phi_J$ with the gauged action $G^{IJ} D^M \Phi_I  D_M \Phi^J$ then the scalar configuration $\Phi = ce^{i\alpha r}$ with zero gauge field can be transformed to $\Phi = c$ with $A = \alpha dr$.\footnote{The field configuration is invariant under the scale transformation ($r \to r + a$), but not invariant under the isometry corresponding to the special conformal transformation.} It is obvious that the former description gives non-zero beta function with cyclic renormalization group (with respect to the coupling constant dual to $\Phi$) while the later description gives the existence of the non-trivial Virial current $J^\mu$ which is dual to the vector field $A$.

Now let us go back to the holographic c-theorem with this additional gauging. The gauging modifies the holographic renormalization group equation as 
\begin{align}
T^{r}_{r} - T^{t}_t = G^{IJ} D_r \Phi_I D_r \Phi_J \ . 
\end{align}
There is a natural counterpart of this equation proposed by Osborn \cite{Osborn:1991gm}. We replace the $\beta$-functions with the gauge invariant $B$-functions:
\begin{align}
\frac{d\tilde{a}}{dr} = G^{IJ} B_I B_J \ . \label{strongestB}
\end{align}
With the operator identity $\beta^I O_I = \partial^\mu J_\mu$, the beta function is ambiguous 
\begin{align}
\beta^I\ &\to \beta^I- (\omega^g g)^I \cr
S &\to S - \omega  \ , \label{gauge}
\end{align}
so we have introduced the gauge invariant $B^I$ function
\begin{align}
B^I &= \beta^I - (S^g g)^I \ . 
\end{align}
This corresponds to the ambiguity whether we do the anti-symmetric wave-function renormalization along the renormalization group flow \cite{Fortin:2012ic}.
Since the central charge is gauge invariant, it is expected that the right hand side is given by the gauge invariant quantities.

With this modified renormalization group consistency equation, the scale invariant but non-conformal field theory $\beta \sim 0$ (but $B \neq 0$) is only possible when the Zamolodchikov metric $G^{IJ}$ is degenerated. In gravity side, this means that when the central charge takes a constant value, we have either $D_r \Phi_I =0$ (conformal invariance) or $G^{IJ}=0$ (degenerate Zamolodchikov metric). It seems unacceptable that we have the degenerate Zamolodchikov metric here because it would violate the unitarity.

Let us comment on some recent results and claims in the literature. In \cite{Fortin:2012ic}, the possibility to obtain scale invariant but non-conformal field theories was attributed to the modified strongest c-theorem: 
\begin{align}
\frac{da}{dr} = G^{IJ} \beta_I B_J \ 
\end{align}
in our convention and it was claimed that $\beta \sim 0$ while $B \neq 0$ is possible (again in our convention). Note however that this does not appear to be the equation indicated by the holographic argument. In our viewpoint, the possibility exists only if $G^{IJ}$ is degenerated.

More recently, \cite{Luty:2012ww} attempted to argue for the non-existence of scale invariant but non-conformal field theory by using the dilaton effective action inspired by  the work \cite{Komargodski:2011vj}, which proved the weak a-theorem in $(1+3)$ dimension. The final results obtained in \cite{Luty:2012ww} seem no stronger than \eqref{strongestB}. It is still an open question whether the right hand side can vanish with non-zero $B$-function. This is again possible when $G^{IJ}$ becomes degenerated and \cite{Luty:2012ww} did not exclude the possibility. In their language, the intermediate channel can interfere and cancel with each other.
In perturbation theory, however, $G^{IJ}$ is positive definite as is well known, so the argument in \cite{Luty:2012ww} is in complete agreement with \eqref{strongestB} as well as in holography.

Indeed, the holographic realization of scale invariant but non-conformal invariant field theories \cite{Nakayama:2009qu}\cite{Nakayama:2009fe}\cite{Nakayama:2011zw} did utilize this possibility of effective vanishing of $G^{IJ}$. Of course, it violates (strict) null-energy condition (or general coordinate invariance), so the physical consistency as a quantum theory of gravity should be considered more carefully.
Finally, all these discussions suggest that foliation preserving diffeomorphic theory of gravity effectively violates the null energy-condition. It will be interesting to pursue the role (and its consistency) of space-time diffeomorphism and violation of null-energy condition further in detail.

\end{document}